\documentclass[prl,twocolumn,letter, showpacs]{revtex4}
\usepackage{amssymb}
\usepackage{amsmath}
\usepackage{bbm}
\usepackage{graphicx}

\begin{document}

\author{Philipp Werner}
\author{Andrew J. Millis}
\affiliation{Columbia University, 538 West, 120th Street, New York, NY 10027, USA}
\title{High-spin to low-spin and orbital polarization transitions in multiorbital Mott systems}

\date{June 30, 2007}

\hyphenation{}

\begin{abstract}

We study the interplay of crystal field splitting and Hund coupling in a two-orbital model 
which captures the essential physics of systems  with two electrons or holes in the $e_g$ shell.
We use 
single site dynamical mean field theory with a recently developed impurity solver which
is able to access strong couplings and low temperatures. 
The fillings of the orbitals and the location of phase boundaries are computed as a function of Coulomb repulsion, exchange coupling and crystal field splitting. We find that the Hund coupling can drive the system into a novel Mott insulating phase with 
vanishing orbital susceptibility. 
Away from half-filling, the crystal field splitting can induce an orbital selective Mott state.

\end{abstract}


\maketitle

The Mott metal-insulator transition plays a fundamental role in electronic condensed matter physics \cite{Imada98}. Much attention has focused on the one-orbital case, in part because of its presumed relevance to high temperature superconductivity \cite{Anderson86} and in part because appropriate theoretical tools for the multiorbital case have until recently not been available. 
In most Mott systems, however, more than one orbital is relevant \cite{Tokura00} 
and the redistribution of electrons among different orbitals leads to new phenomena such as orbital ordering or ``orbital selective" Mott transitions. Recent studies of nickelates \cite{Kunes}, titanates \cite{Ulrich06}, cobaltates \cite{Ishida05}, manganates \cite{Yamasaki06}, vanadates \cite{Biermann05, Lechermann05, Yoshida05} and ruthenates \cite{Tokura00, Koga04, Okamoto04, Liebsch07} have focused interest on the interplay between the Mott metal-insulator transition and orbital degeneracy.
A fundamental question in this field, relevant in particular to the issue of lattice distortions in strongly correlated materials, is the response of multi-orbital systems to a perturbation which breaks the orbital degeneracy. In this paper, we show that a two orbital model with Hund coupling and crystal field splitting exhibits two fundamentally different Mott phases, one characterized by a vanishing orbital susceptibility, and one adiabatically connected to the band insulating state. We characterize these phases in terms of the atomic ground states. 

Multiorbital models are more difficult to study both because of the larger number of degrees of freedom, and because the physically important exchange and pair-hopping terms are not easy to treat by standard Hubbard-Stratonovich methods \cite{Sakai06}. Weak coupling approaches \cite{Okamoto04} have been used to show that exchange and pair hopping interactions act to suppress the response to a crystal field splitting, and some authors have studied the model without exchange and pair hopping terms \cite{Lechermann05}, but a reliable extension of these results to physically relevant Slater-Kanamori interactions and the strong coupling regime has been lacking. 

Dynamical mean field theory (DMFT) provides a non-perturbative and computationally tractable 
framework to study correlation effects and has allowed insights into the Mott 
 metal-insulator transition \cite{Georges96}. In its single site version, DMFT ignores the momentum dependence of the self-energy and reduces the original lattice problem to the self-consistent solution of a quantum impurity model given by the Hamiltonian
$H_\text{QI} = H_\text{loc}+H_\text{hyb}+H_\text{bath}$.
For multi-orbital models $H_\text{loc} = \sum_{m}\epsilon_m c^\dagger_{m} c_{m} + \sum_{j,k,l,m}U^{jklm}c^\dagger_j c_k c^\dagger_l c_m$, where $m=(i,\sigma)$ denotes both orbital and spin indices, and $U^{jklm}$ some general four-fermion interaction. $H_\text{hyb}$ and $H_\text{bath}$ are the impurity-bath mixing and bath Hamiltonians, respectively.
While the DMFT approximation simplifies the problem enormously (replacing a $3+1$ dimensional field theory by a quantum impurity model plus a self consistency condition),  the extra complications associated with exchange couplings in multiorbital systems have until recently prohibited extensive numerical work.
Interesting progress has been made using a finite temperature exact diagonalization technique \cite{Ishida05, Liebsch07}, but this approach requires a truncation of $H_\text{bath}$ to a small number of levels.  In Refs.~\cite{Werner05, Werner06} we have introduced a continuous-time impurity solver which can handle the general interactions in $H_\text{loc}$. The method, which is free from systematic errors, is based on a diagrammatic expansion of the partition function in the impurity-bath hybridization $H_\text{hyb}$. 
 
Here, we employ this solver to study the physically relevant case in which the number of electrons matches the number of orbitals. 
The local Hamiltonian is
\begin{align}
&H_\text{loc}=-\sum_{\alpha=1,2}\sum_{\sigma}\mu n_{\alpha,\sigma}+ \sum_\sigma\Delta(n_{1,\sigma}-n_{2,\sigma})\nonumber\\
&\hspace{3mm}+\sum_{\alpha=1,2} U n_{\alpha,\uparrow} n_{\alpha,\downarrow}+\sum_\sigma U' n_{1,\sigma} n_{2,-\sigma} \nonumber\\
&\hspace{3mm}+ \sum_\sigma (U'-J) n_{1,\sigma}n_{2,\sigma}\nonumber\\
&\hspace{3mm}-J(\psi^\dagger_{1,\downarrow}\psi^\dagger_{2,\uparrow}\psi_{2,\downarrow}\psi_{1,\uparrow}
+ \psi^\dagger_{2,\uparrow}\psi^\dagger_{2,\downarrow}\psi_{1,\uparrow}\psi_{1,\downarrow} + h.c.),
\label{H_delta}
\end{align}
with $\mu$ the chemical potential, $\Delta$ the crystal field splitting, $U$ the intra-orbital and $U'$ the inter-orbital Coulomb interaction, and $J$ the coefficient of the Hund coupling. We adopt the conventional choice of parameters,  $U'=U-2J$, which follows from symmetry considerations for $d$-orbitals in free space and is
also assumed to hold in solids. With this choice the Hamiltonian (\ref{H_delta}) is rotationally invariant in orbital space and the condition for half-filling becomes $\mu =\mu_{1/2}\equiv \frac{3}{2}U-\frac{5}{2}J$. In the DMFT self-consistency loop we use a semi-circular density of states of bandwith $4t$ (Bethe lattice). 
The temperature, unless otherwise noted, is $T/t=0.02$ and we suppress magnetic order by averaging over spin up and down in each orbital. No sign problem is encountered in the simulations. 

\begin{figure}[t]
\begin{center}
\includegraphics[angle=-90, width=0.9\columnwidth]{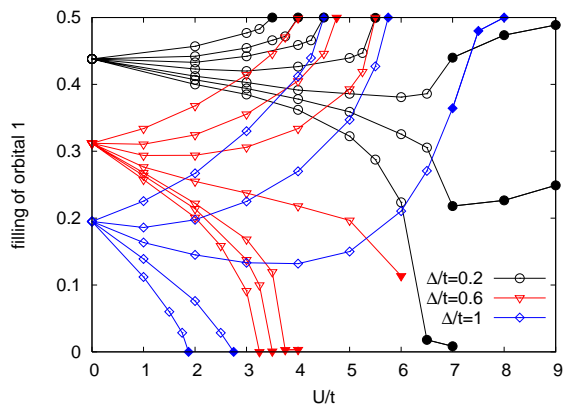}
\caption{Filling of orbital 1 as a function of $U$ for $\Delta/t=0.2$, 0.6, 1 and several values of $J/U$. The different curves for given $\Delta$ correspond (from bottom to top) to $J/U=0$, (0.01), (0.02), 0.05, 0.1, 0.15, 0.25, respectively. Open (full) symbols correspond to metallic (insulating) solutions. The metal-insulator transition 
is characterized by a jump in filling and a coexistence region where both insulating and metallic solutions exist. Our data show the region of stability of the metallic phase.  
}
\label{filling_u_delta}
\end{center}
\end{figure}

\begin{figure}[t]
\begin{center}
\includegraphics[angle=-90, width=0.9\columnwidth]{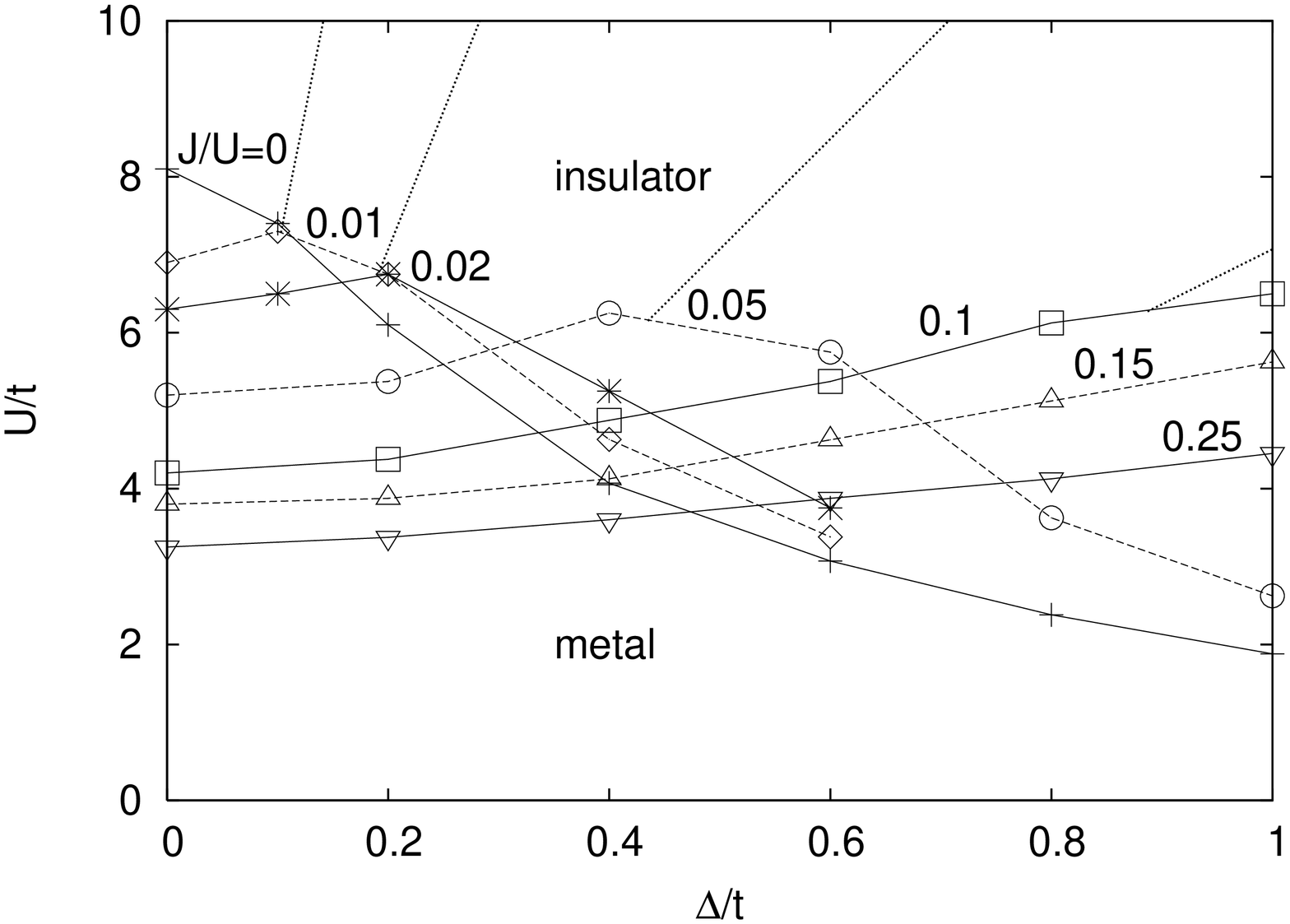}
\includegraphics[angle=-90, width=0.9\columnwidth]{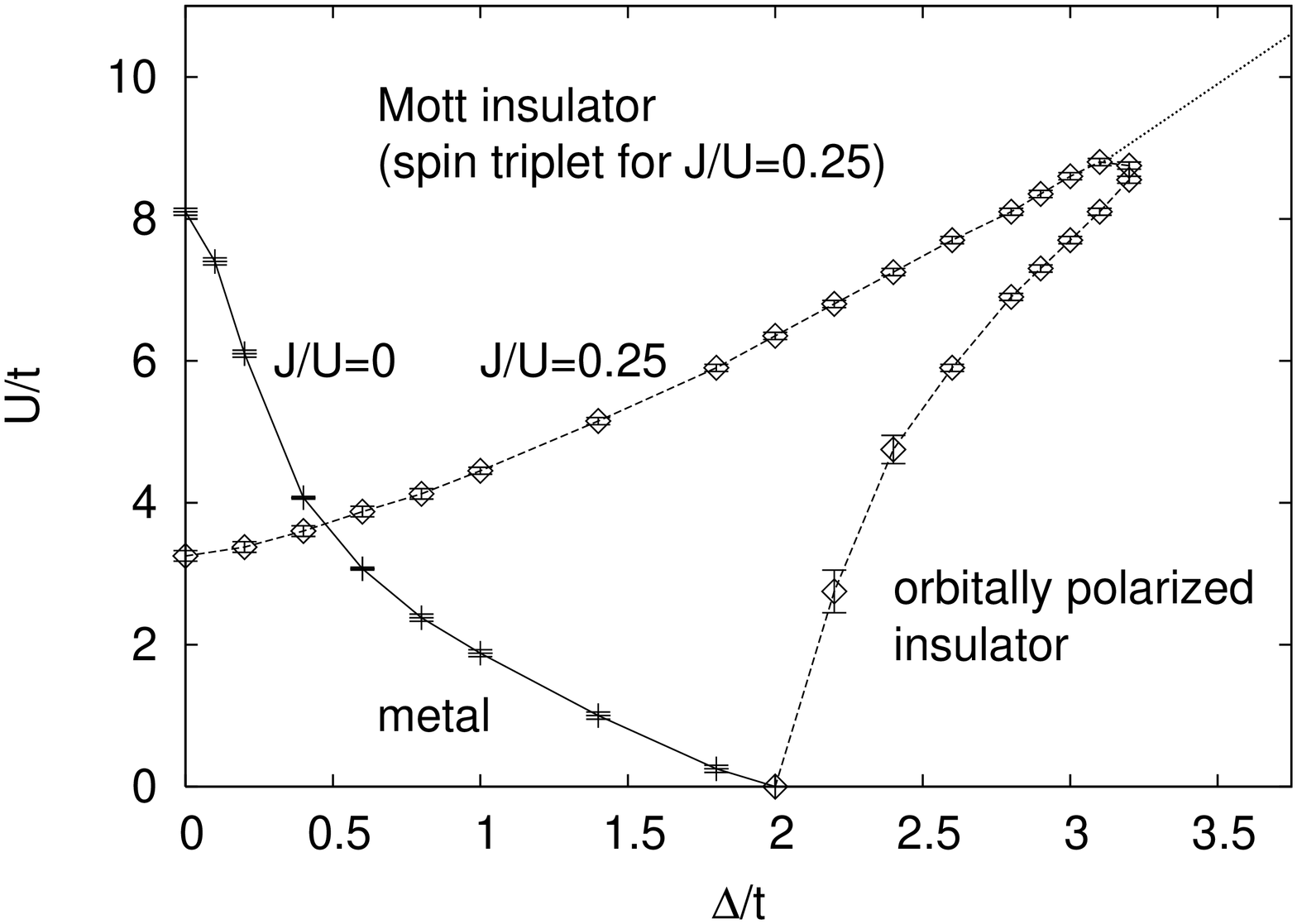}
\caption{Phase diagram in the plane of crystal field splitting $\Delta$ and intraorbital Coulomb repulsion $U$ for indicated values of $J/U$. For $J=0$ the phase boundary is a monotonic function of $\Delta$, whereas for $J/U>0$ it peaks near $\Delta = \sqrt{2}J$ (indicated by the dotted lines). The insulating state in the region $\Delta \lesssim \sqrt{2} J$ is characterized by a vanishing orbital susceptibility. 
}
\label{u_delta}
\end{center}
\end{figure}

The main result is shown in Fig.~\ref{filling_u_delta}, which for several values of $\Delta$ and $J/U$ plots the filling per spin, $n_1$, of orbital 1 as a function of interaction strength. The half filling, non-magnetic condition implies $n_2=1-n_1$. In the $T\rightarrow 0$ limit, three phases are found: a metallic phase (which may
have any value of $n_1$ between $0$ and $0.5$), an orbitally polarized insulator 
favored by large $\Delta$ and small $J$, and a Mott insulator (with $n_1=0.5=n_2$)
favored by large $U$, small $\Delta$ and large $J$. 
If $U$ 
is increased from zero to a small value, the orbital splitting either 
increases (small $J/U$) or decreases (large $J/U$), consistent
with the findings of Ref.~\cite{Okamoto04}.  
As interaction strength is further increased, one of several things may happen: at very small $J/U$, $n_1$ continues to decrease, and the system eventually undergoes a transition to an orbitally polarized insulator (for large $\Delta$ essentially a band insulator). For somewhat larger $J/U$, the occupancy $n_1$, after an initial decrease, goes through a minimum and begins to increase. At even stronger interactions, one then observes a transition either to an orbitally polarized insulator (where $n_1$ may take a range of values) or into a special type of insulator with $n_1=0.5$. 

Figure~\ref{u_delta} shows the 
metal-insulator phase diagram in the space of crystal field splitting and 
Coulomb repulsion for several values of $J/U$.
In the absence of a crystal field splitting ($\Delta=0$), we observe a metal-insulator transition at a strongly $J$-dependent critical $U$. This finding is consistent with data presented in Ref.~\cite{Koga02}. 
As $\Delta$ is increased, the critical $U$ changes. For $J=0$ and fixed $U$, $n_1$ decreases until the band is emptied and a metal-insulator transition occurs. The monotonic decrease of the critical $U$ with $\Delta$ at $J=0$ is a special case. For $J>0$, the first effect of a small $\Delta$ is to stabilize the metallic phase. Then, at larger $\Delta$, a reentrant insulating phase occurs. We shall show below that this behavior arises from the unusual nature of the insulating state at $J>0$ and small $\Delta$, which is characterized at $T=0$ by a strictly vanishing orbital susceptibility. If $\Delta$ is increased at large $U$, this state makes a transition to an orbitally polarized insulator at $\Delta \approx \sqrt{2}J$. We therefore plot in Fig.~\ref{u_delta} the curves $\Delta=\sqrt{2}J$ as dotted lines, and suggest that they correspond to the $T=0$ phase boundary between two distinct insulating states. 

\begin{figure}[t]
\begin{center}
\includegraphics[angle=-90, width=0.9\columnwidth]{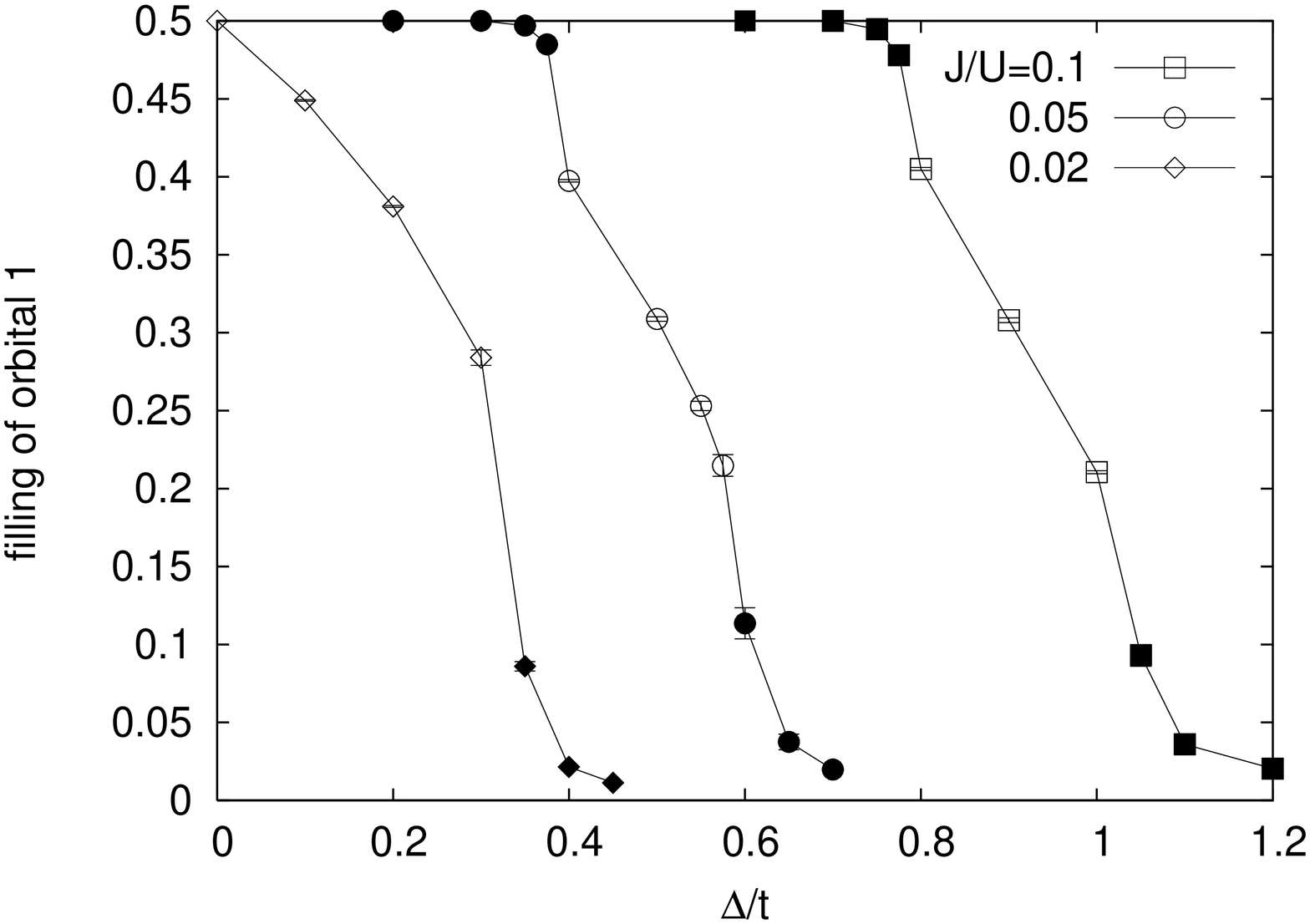}
\includegraphics[angle=-90, width=0.9\columnwidth]{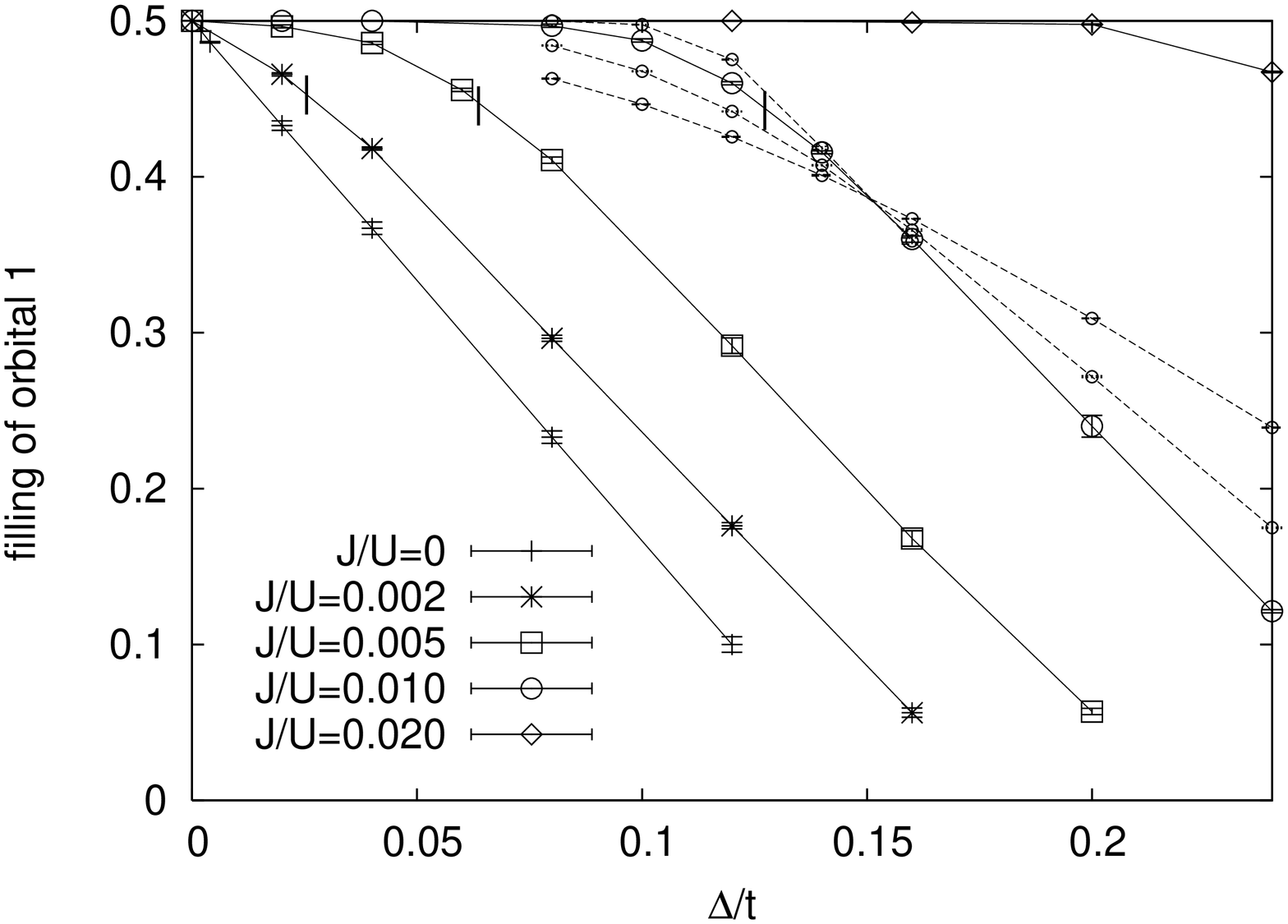}
\caption{Filling of orbital 1 as a function of $\Delta$ for fixed $U$ and indicated values of $J/U$.
Top panel: $U/t=6$. Open (full) symbols correspond to metallic (insulating) solutions. Bottom panel: $U/t=9$. Here, all solutions are insulating. For crystal field splittings smaller than $\Delta_c=\sqrt{2}J$ (indicated by a vertical line) the orbital susceptibility in the $T\rightarrow 0$ limit is completely suppressed. Solid lines are for $\beta t=50$, dotted lines show results for $\beta t=12.5$, 25 and 100, respectively.
}
\label{j}
\end{center}
\end{figure}

Figure~\ref{j} plots the filling of orbital 1 as a function of crystal field splitting for fixed $U/t$ and several values of $J/U$. 
The leftmost curve in the upper panel shows the density variation for $J/U=0.02$. At $\Delta=0$, the model is metallic. The rapid variation of $n_1$ with $\Delta$ reflects the large, but finite orbital susceptibility of the metal, which for this small value of $J$ is strongly enhanced by $U$. At $\Delta/t\approx 0.325>\sqrt{2}J$, an apparently first order transition occurs to the orbitally polarized insulating state, which then evolves smoothly (as $\Delta$ is increased) to the band insulator ($n_1\rightarrow 0$).
The two larger $J$ values reveal a different behavior. For $\Delta<\Delta_c\approx \sqrt{2} J$ the insulating state is characterized by an orbital occupancy which is independent of crystal field splitting. Then, an apparently discontinuous transition occurs to a metallic state with a large orbital susceptibility, which at even larger $\Delta$ exhibits a first order transition to the orbitally polarized insulating state. The lower panel of Fig.~\ref{j} shows the behavior for larger $U$, where the model is always insulating. Our data for $J=0$ exhibit a rapid variation of $n_1$ with $\Delta$. The slope is set by the inverse of the Kugel-Khomskii superexchange $\sim t^2/U$; thermal effects are unimportant at $\beta t =50$. 
For $J>0$ and small $\Delta$, the model is insulating, with a vanishing orbital susceptibility, then (near $\Delta=\sqrt{2}J$) makes a transition to the orbitally polarized insulating phase with a differential susceptibility $\partial n_1/\partial\Delta$ determined by Kugel-Khomskii physics. 
Note that the transition between the two insulators is sharp only at $T=0$; for $T>0$ a rapid (but smooth) crossover occurs. 

To gain insight into these phenomena, we look at the contribution to the partition function from the different eigenstates of the local Hamiltonian. $H_\text{loc}$ has 16 eigenstates, which we number essentially as in Table~II of Ref.~\cite{Werner06}. For the following discussion it is important to note that  
$|6\rangle$, $|7\rangle$ and $|8\rangle$ are the three spin triplet states (with energy $U-3J-2\mu$), while $|10\rangle$ and $|11\rangle$ are linear combinations of  the states $|\!\!\uparrow\downarrow, 0\rangle$ and $|0,\!\uparrow\downarrow\rangle$ with two electrons in one orbital and none in the other. 
The latter two states are coupled by the pair hopping and affected by the crystal field splitting. Here, we choose them to be eigenstates of $H_\text{loc}$ corresponding to the eigenenergies $U\pm \sqrt{J^2+4\Delta^2}-2\mu$: $(1+\alpha_\pm^2)^{-1/2}(|\!\uparrow\downarrow, 0\rangle+\alpha_\pm|0,\!\uparrow\downarrow\rangle)$, $\alpha_\pm=\pm(\sqrt{J^2+4\Delta^2}\mp2\Delta)/J$.  
In particular, we choose $|10\rangle$ to be the eigenstate with lower energy.

\begin{figure}[t]
\begin{center}
\includegraphics[angle=-90, width=0.9\columnwidth]{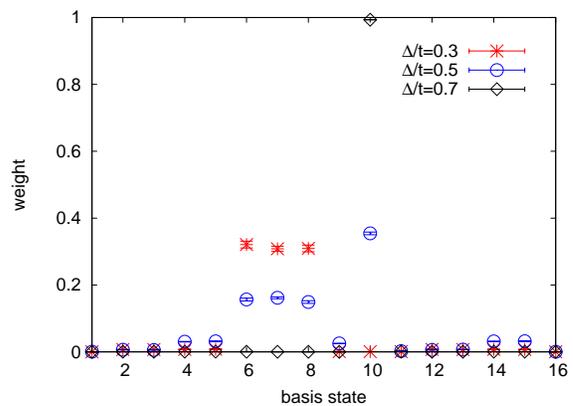}
\caption{
Weight of the different eigenstates of $H_\text{loc}$ for $U/t=6$, $J/U=0.05$ and $\Delta/t=0.3$, 0.5 and 0.7. The smallest crystal field splitting corresponds to an insulating state with suppressed orbital susceptibility, the intermediate value to a metallic state and the largest splitting to a ``band insulator" (see Fig.~\ref{j}). 
}
\label{weight}
\end{center}
\end{figure}

Figure~\ref{weight} shows the weights of these states for the three phases found at $U/t=6$, $J/U=0.05$ (see Fig.~\ref{j}). In the small-$\Delta$ phase, the triplet states are occupied, with small excursions into states with occupancy 1 or 3. We therefore call this phase the triplet Mott insulator. 
The triplet states of course have one electron in each orbital and gain no energy from orbital polarization (the remarkable fact is that this feature is preserved after coupling to the lattice).
In the metallic phase, a large number of states is visited, while in the orbitally polarized insulator, the dominant local state (whose weight increases continuously with $\Delta$) is a singlet ($|10\rangle$). The triplet states are almost completely suppressed in the orbitally polarized phase.  The large-$U$ insulator-insulator transition exhibits the same features, but without the intermediate metallic phase, and is therefore also a transition between high and low spin states.  
Comparison of the eigenenergies of the spin triplet states and $|10\rangle$ show that these levels cross at $\Delta = \sqrt{2}J$. Thus, the transition from triplet Mott insulator to orbitally polarized insulator occurs at $\Delta_c=\sqrt{2}J$, consistent with our numerical data. We also note that the wave-function of state $|10\rangle$ depends on the ratio $J/\Delta$, leading in the large-$\Delta$ limit to $n_1(\Delta)\approx (J/4\Delta)^2$. In the low spin phase, the orbital susceptibility has therefore two contributions: one originating from Kugel-Khomskii physics 
and one of order $J^2/\Delta^3$ from $H_\text{loc}$. The latter explains the roundings seen in the right most curve of Fig.~\ref{j}.



\begin{figure}[t]
\begin{center}
\includegraphics[angle=-90, width=0.9\columnwidth]{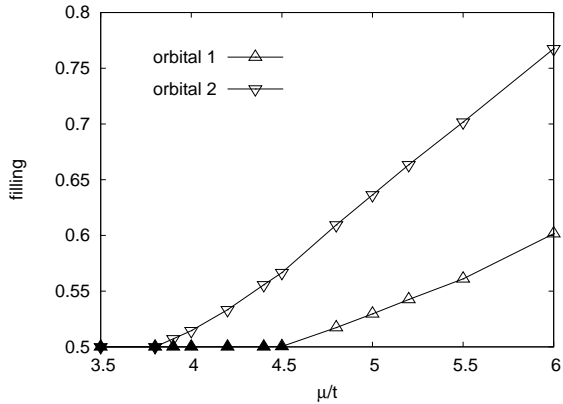}
\caption{Filling $n_1(\mu)$, $n_2(\mu)$ for $U/t=4$, $J/U=0.25$ and $\Delta/t=0.4$. Full (open) symbols correspond to insulating (metallic) solutions. At half-filling ($\mu/t=3.5$), the system is in a triplet Mott insulating state, for $3.9\lesssim \mu/t \lesssim 4.6$ in an orbital selective Mott state, and for $\mu/t\gtrsim 4.6$ metallic in both bands. }
\label{orbital_selective}
\end{center}
\end{figure}

We briefly address the issue of the orbital selective Mott transition, which provides a mechanism for local moment formation in correlated materials, and has been the subject of much recent debate \cite{Koga04}.  Previous studies focused on two-orbital models with different band-widths and integer number of electrons. We find that in the presence of a crystal field splitting, shifting the chemical potential can drive the system into an orbital selective Mott state, even if the band-widths are equal. Figure~\ref{orbital_selective} shows the filling per spin in both orbitals as a function of $\mu$, for $U/t=4$, $J/U=0.25$ and $\Delta/t=0.4$. Doping occurs first in one of the bands, leaving the other in a Mott state with   a magnetic moment. Further change of the chemical potential drives the second band metallic. 


In conclusion, we have shown that multiorbital impurity models with
realistic couplings can be efficiently simulated with the method
of Ref.~\cite{Werner06}.
We have presented numerical evidence, based on single site DMFT calculations, for the existence of two distinct Mott insulating phases in a half-filled two-orbital model with Hund coupling and crystal field splitting. At strong interactions and $J<\sqrt{2}\Delta$, the system, in the $T\rightarrow 0$ limit, is
in  a phase  characterized by a vanishing orbital susceptibility, and a spin
$1$ moment on each site.
For $J>\sqrt{2}\Delta$ an orbitally polarized insulator 
is found.  
The  exchange terms promote insulating behavior at $\Delta=0$ but can  
stabilize a metallic phase at values of $\Delta$ for which the non-interacting model is a band-insulator. 

It is interesting to compare our results to recent work on the bilayer Hubbard model \cite{Fuhrmann06, Kancharla07}. The model which these authors study is equivalent to our model with $U=U'=J$, and $\Delta$ replaced by the interlayer hopping. In the low energy sector of this model, only four states (essentially our three triplets and the pair hopping state $|10\rangle$) are relevant, and what these authors describe as the Mott insulator to band insulator crossover corresponds to our transition (apparently sharp at $T=0$) between triplet Mott insulator and orbitally polarized insulator.


The existence of two distinct insulating phases raises many interesting questions
including the theory of an insulator with strictly vanishing orbital susceptibility (which should exhibit an orbital gauge symmetry)
and the nature and properties of the different metal-insulator transitions. The physics near the ``triple point" remains to be studied.
Our results away from half-filling 
suggest that lightly doped La$_2$NiO$_4$ is in an orbitally selective Mott phase. 

\acknowledgements

The calculations have been performed on the Hreidar beowulf cluster at ETH Z\"urich, using the ALPS-library \cite{ALPS}. We thank M. Troyer for the generous allocation of computer time, A.~Georges and A.~Poteryaev for stimulating discussions and 
NSF-DMR-040135 for support.

\end{document}